\documentclass[aps,prd,twocolumn,a4paper,final,superscriptaddress,longbibliography]{revtex4-1}

\usepackage{graphicx}
\usepackage{hyperref}

\usepackage [latin1]{inputenc}

\usepackage{amsmath,amssymb}
\usepackage{bm} 
\usepackage{mathtools}
\usepackage[caption=false]{subfig}
\usepackage{booktabs}
\usepackage{mathrsfs}

\usepackage{cases}

\usepackage{color}
\usepackage{ulem}
\usepackage{comment}
\usepackage{cleveref}

\usepackage{natbib}

\graphicspath{{./Fig/}}

\begin{document}

\title{A baby--Skyrme model with anisotropic DM interaction: Compact skyrmions revisited}

\author{Funa Hanada}
\email{funa87@gmail.com}

\author{Nobuyuki Sawado}
\email{sawadoph@rs.tus.ac.jp}
\affiliation{Department of Physics and Astronomy, Tokyo University of Science, Noda, Chiba 278-8510, Japan}

\begin{abstract}
We consider a baby--Skyrme model with Dzyaloshinskii--Moriya interaction (DMI) and two types of potential terms. 
The model has a close connection with the vacuum functional of fermions coupled with $O(3)$ nonlinear 
$\bm{n}$-fields and with a constant $SU(2)$ gauge background. 
The energy functional is derived from the heat-kernel expansion for the fermion determinant. 
The model possesses normal skyrmions with topological charge $Q = 1$. 
The restricted version of the model also includes both the weak-compacton case
(at the boundary, not continuously differentiable) and genuine-compacton case (continuously differentiable).
The model consists of only the Skyrme term, and 
the DMI provides soliton solutions that are known as \textit{skyrmions without any potential}.
The BPS equation in the supersymmetric soliton models implies that the impurity coupling 
is closely related to the DMI. 
Therefore, the effect of an exponentially localized DMI is also studied in the present model. 

\end{abstract}

\maketitle

\section{Introduction}

The Skyrme model, a (3+1)-dimensional nonlinear field theory of pions, is a model of hadrons and is 
supposedly the most promising and long-lived effective model in the low-energy domain of 
quantum-chromodynamics (QCD). The skyrmions, the topological solitons in the Skyrme model, suitably describe 
not only the standard hadrons and nuclei but also structures of the dense nuclear matter
~\cite{Adam:2010fg,Adam:2013tda,Adam:2015lra,Ferreira:2021ryf} and the neutron star~\cite{Adam:2014dqa,Adam:2015lpa}. 

The Skyrme model in (2+1)-dimensions has recently gained considerable attention.
Particularly, magnetic skyrmions have garnered increasing interest in both theoretical aspects of topological matter 
and also in many applications of spintronics, quantum computing, and dense magnetic nanodevices. 
Magnetic skyrmions are derived from a model encompassing Dzyaloshinskii--Moriya interaction (DMI)
~\cite{DZYALOSHINSKY1958241,Moriya:1960}. 
The DMI and a potential break in the scale invariance of the model successfully evade Derrick's theorem. 
The Skyrme field $\bm{n}=(n_1,n_2,n_3)$ with $\bm{n}\cdot\bm{n}=1$, realizes maps: $S^2\to S^2$, 
and are characterized by the homotopy group $\Pi_2(S^2)=\mathbb{Z}$.
The energy density is defined as~\cite{Bogdanov:1989,BOGDANOV:1994, Barton-Singer:2018dlh,Schroers:2019hhe}
\begin{align}
&\mathcal{E}_{\textrm{DM}}=\kappa_2(\partial_i \bm{n} )^2 +\kappa_1\bm{n}\cdot \nabla\times \bm{n} +V[\bm{n}]\,, ~~i=1,2,
\label{magskyrme}
\end{align}
where $\kappa_2,\kappa_1$ are constants with a positive sign. 
The second differential term (the kinetic term) is a scale-invariant term, and the DMI has a negative contribution 
to the energy; accordingly, the solution may exist in terms of Derrick's theorem. 

The baby--Skyrme model is a direct replica of the (3+1)--Skyrme model, and
the model consists of an $O(3)$ nonlinear sigma model (the kinetic term), 
a 4th-order differential term (the Skyrme term) and a Zeeman or other types of potential terms.
As is widely known that the Skyrme and the potential terms are responsible for Derrick's theorem, the energy density of the baby--Skyrme model is defined by~\cite{Piette:1994ug} 
\begin{align}
&\mathcal{E}_{\textrm{bS}}=\kappa_2(\partial_i \bm{n} )^2 +\kappa_4\left(\partial_i \bm{n } \times \partial_j \bm{n }\right)^2 +V[\bm{n}]\,,
~~i,j=1,2,
\label{babyskyrme}
\end{align}
where $\kappa_4$ is a positive constant. 
The baby-skyrmions have applications in terms of quantum Hall effects
~\cite{Sondhi:1993,Neubauer:2009_1,Neubauer:2009_2,Balram:2015,Jiang:2017}, 
nematic crystals~\cite{Bogdanov:2003,Fukuda:2011,Leonov:2014,Ackerman:2017,Matteis:2018,Matteis:2022}, 
superconducting materials~\cite{Zyuzin:2017}, 
and brane-world scenarios~\cite{Kodama:2008xm,Brihaye:2010nf,Delsate:2011aa,Delsate:2012hz}, so on. 
The baby--Skyrme model without the kinetic term, named 
the restricted baby--Skyrme model~\cite{Gisiger:1996vb,Adam:2010jr,Andrade:2022wiv}, 
has a significant feature: it possesses analytical Bogomol'nyi--Prasad--Sommerfield (BPS) solutions. 
The baby--Skyrme model and the restricted model provide solutions pertaining to 
compacton. Compactons possess a distinct character among other solutions of standard field theory models. 
The field considers its vacuum values outside this support, and the energy and charge are always
concentrated on the compact support~\cite{Arodz:2005gz,Arodz:2008jk}. 
There have been several studies of compact skyrmions in the baby--Skyrme model
~\cite{Gisiger:1996vb,Adam:2009px, Speight:2010sy, Adam:2010jr,Ashcroft:2015jwa,Casana:2022bei}. 
For determining compactons, the baby--Skyrme model requires a non-analytical potential called V-shaped potential.
While in the restricted model, other choices for the potential may be available, a prominent challenge to the modification of the model exists. 
The baby--Skyrme model with fractional power of the kinetic term with no potential term successfully evades 
Derrick's theorem and has compact and non-compact skyrmion solutions~\cite{Ashcroft:2015jwa}.

A natural question arises here: Can both models be combined to describe the phenomenology? 
At this point, we have no clear evidence that both interactions should coexist. 
However, from a theoretical perspective, 
it may be effective to consider a combined model and find novel solutions. 
In this study, we examine such models and find several types of solutions, including compactons. 
For simplicity, we focus on the rotationally symmetric solutions; 
however, if the constraint is lifted, various structures will emerge. 

In \cite{Adam:2019yst}, the authors studied the supersymmetric extensions of a restricted baby--Skyrme 
model of the squared Zeeman potential with ``the impurity coupling. 
In particular, the analytical solution in the BPS equation is found for the exponentially localized 
impurity $\sim e^{-\beta r}$.  Therefore, it is worth to investigate the present model with the DMI of 
the exponentially localized impurity. 

The paper is organized as follows. In Section \ref{sec:2} we introduce a fermionic model and the 
resulting topological information from the imaginary part of the action. 
Also a brief explanation of our model, including the energy functional and the Euler equation is
done in this section. 
We present several analytical and numerical solutions to the model in Section \ref{sec:3}. 
We describe a novel combined model that has no potential term and provides the solutions in Section \ref{sec:4}.
In Section \ref{sec:5} we discuss the impurity models, in which we consider the exponentially localizing 
DMI coupling in the model.  
The conclusions and remarks are presented in the last section.

\section{\label{sec:2}The model}

In this paper, we analyze a Skyrme--type model with the DMI. The energy is defined as
\begin{align}
    E & = \int d^2 x \Bigl\{\kappa_2 \left(\partial_i \bm{n} \right)^2 
\nonumber \\
    &+ \kappa_1 \bm{n} \cdot \left(\nabla \times \bm{n} \right)
    -\kappa_1 (\partial_1n_2-\partial_2n_1)
\nonumber \\
    & + \kappa_{4a} \left(\partial_i \bm{n} \times \partial_j \bm{n} \right)^2 + \kappa_{4b} \left(\partial_i \bm{n} \right)^2 \left(\partial_j \bm{n} \right)^2 
\nonumber\\
    & +\kappa_{0a}\left(1-n_3 \right)+\kappa_{0b}\left(1-n_3 \right)^2 \Bigr\}\,,
    \label{eq:edens}
\end{align}
where each term corresponds to
\begin{align}
&\textrm{(i) the kinetic}:~
E_2:=\kappa_2\int d^2 x \left(\partial_i \bm{n} \right)^2\,,
\nonumber \\
&\textrm{(ii) the DMI}:~
E_1:=\kappa_1\int d^2x \bm{n} \cdot \left(\nabla \times \bm{n} \right)\,,
\nonumber \\
\nonumber \\
&\textrm{(iii) the Skyrme}:~
E_{4a}:=\kappa_{4a}\int d^2x \left(\partial_i \bm{n} \times \partial_j \bm{n} \right)^2\,,
\nonumber \\
&\textrm{(iv) an extended 4th}:~
\nonumber \\
&\hspace{2.cm}E_{4b}:=\kappa_{4b}\int d^2x \left(\partial_i \bm{n} \right)^2 \left(\partial_j \bm{n} \right)^2\,,
\nonumber \\
&\textrm{(v) the Zeeman}:~
E_{0a}:=\kappa_{0a}\int d^2x \left(1-n_3 \right)\,,
\nonumber \\
&\textrm{(vi) a squared Zeeman}:~
E_{0b}:=\kappa_{0b}\int d^2x  \left(1-n_3 \right)^2\,.
\end{align} 
Note that the integration of the vortex strength is zero. 

Although the model might be considered as just a hybrid of the above magnetic Skyrme model~(\ref{magskyrme}) 
and the baby--Skyrme model~(\ref{babyskyrme}), it actually has a systematic origin. 
In~\cite{Jaroszewicz:1985ip,Abanov:2000ea,Abanov:2001iz,Amari:2019tgs}, 
the authors investigated the $O(3)$ nonlinear sigma model Lagrangian and also their topological terms 
based on the derivative expansion of the Lagrangian of the fermions coupled with the Skyrme field 
via $\partial_\mu \bm{n}$. There are certain recent theoretical studies regarding the fermions with 
the baby-skyrmions~\cite{Perapechka:2018yux} and the magnetic skyrmions~\cite{Perapechka:2019upv}, 
considering the backreaction from the fermionic fields. 
We begin with the following vacuum functional:
\begin{align}
\mathcal{Z}= \int \mathscr{D}\psi\mathscr{D}\bar{\psi}e^{S_{\rm E}}
\label{vacuumfunctional}
\end{align}
where the Euclidean action is
\begin{align}
S_{\rm E}=\int d\tau\int d^2x \biggl[\bar{\psi}\Bigl(i\gamma_\mu (\partial_\mu-i\bm{A}_\mu)-m\bm{\tau}\cdot\bm{n}\Bigr)\psi\biggr]\,.
\end{align}
The Euclidean time component $\tau$ is defined by the Wick-rotation $t=x_0=-i\tau$. 
The gamma matrices are defined as $\gamma_\mu:=-i\sigma_\mu, \mu=1,2,3$ that satisfy the 
Clifford algebra $\{\gamma_\mu,\gamma_\nu\}=-2\delta_{\mu\nu}$. 
Although the Yukawa coupling constant $m$ is theoretically free to be chosen, 
the presence of the fermionic zeromodes requires that it be above the critical 
value $m\geqq m_0$~\cite{Amari:2019tgs}. In the following, we regard $m$ as $m\geqq 1$ 
without loss of generality. 
 
The DMI term emerges introducing a constant
background gauge field $\bm{A}_\mu=A_\mu^a\tau_a/2$~\cite{Schroers:2019hhe, Amari} defined as
\begin{align}
&A_1^a=(-D,0,0),~~A_2^a=(0,-D,0),
\nonumber \\
&\hspace{1cm}\textrm{all the others are zero}\,.
\label{DMgauge}
\end{align}
Performing the integration (\ref{vacuumfunctional}), we obtain the effective action $\omega(\bm{n})$
\begin{align}
\mathcal{Z}=\det i\mathcal{D}\equiv \exp[\omega(\bm{n})],~~
\omega(\bm{n}):=\textrm{Tr}\log (i\mathcal{D})\,,
\label{efaction1}
\end{align}
where the Dirac operator is expressed as
\begin{align}
i\mathcal{D}:=i\gamma_\mu (\partial_\mu-i\bm{A}_\mu)-m\bm{\tau}\cdot\bm{n}\,.
\label{Diracop}
\end{align}
In Euclidean space, the effective action is generally a complex quantity 
$\omega(\bm{n}):=\omega_R(\bm{n})+i\omega_I(\bm{n})$, where
\begin{align}
&\omega_R(\bm{n})=\frac{1}{2}\textrm{Tr}\log\mathcal{D}^\dagger\mathcal{D}\,,~~
\label{efactionr} \\
&\omega_I(\bm{n})=\frac{1}{2i}\textrm{Tr}\log(\mathcal{D}^\dagger)^{-1}\mathcal{D}\,.
\label{efactioni}
\end{align}

The real component here needs to be dealt with because it generates an effective Skyrme-type model. 
We perform the expansion based on the heat-kernel method~\cite{Ebert:1985kz, Reinhardt:1989st} 
that directly investigates the static energy of the model. The calculations up to 4th differential
order terms to the energy are almost straightforward; nonetheless, the results are cumbersome. 
For all the 3rd and 4th differential order terms, we therefore set $D= 0$ to simplify the model. 
We describe the detailed analysis in Appendix A. 

The imaginary part of the action (\ref{efaction1}) conveys the statistical property of the model. 
Thus, we consider the $U(1)$ gauged model of (\ref{Diracop})
\begin{align}
i\mathcal{D}:=i\gamma_\mu (\partial_\mu-i\bm{A}_\mu-ia_\mu)-m\bm{\tau}\cdot\bm{n}
\end{align}
where $a_\mu$ is an external electromagnetic potential. After attempting to expand $a_\mu$, such 
that it contributes to the effective action as follows~\cite{Abanov:2000ea,Abanov:2001iz}
\begin{align}
\omega_I(\bm{n})=-\int d^3x a_\mu J_\mu
\end{align}
where the topological current is
\begin{align}
&J_\mu=\frac{1}{16\pi i}\epsilon_{\mu\nu\delta}\textrm{tr}(u D_\nu uD_\delta u)\,,
\\
&D_\mu u:=\partial_\mu u-i[A_\mu,u],~~~~u:=\bm{\tau}\cdot\bm{n}\,,
\end{align}
and the third component becomes 
\begin{align}
J_3=\frac{1}{4\pi}\biggl(\epsilon_{abc}n_a\partial_1n_b\partial_2n_c
+D(\partial_1n_2-\partial_2n_1)+D^2n_3\biggr)\,.
\label{topologicaldensity}
\end{align}
The first term defines the well-known topological charge 
\begin{align}
Q&=\frac{1}{4\pi}\int d^2 x q(\bm{x})\nonumber\\
    &=\frac{1}{4\pi}\int d^2 x \bm{n}(\bm{x}) \cdot \left\{\partial_1 \bm{n}(\bm{x}) \times \partial_2 \bm{n}(\bm{x})\right\}\,.
\end{align}

The configuration space of the model comprises maps from the plane $\mathbb{R}^2$ to the target space $S^2$. 
Considering coordinates $\Theta, \Phi$ on the target sphere (corresponding to the usual spherical polar coordinates), 
the best-known solution is the rotationally symmetric solution expressed as 
\begin{align}
    \Theta =f(r),~~\Phi=\varphi\,,
\end{align}
where $r,\varphi$ are the usual polar coordinates on the plane. 
Consequently, the configuration giving rise to a baby-skyrmion with topological charge $n$ is defined by
\begin{align}
    \bm{n}=\left(\sin f(r) \cos \left(n \varphi+\gamma\right), \sin f(r) \sin  \left(n \varphi+\gamma\right), \cos f(r) \right)\,,
    \label{eq:ansatz}
\end{align}
where $n\in \mathbb{N}$ and the phase $\gamma$ describes the internal orientation of the solution. 
Imposing the boundary condition 
\begin{align}
f(0)=\pi,~~f(\infty)=0\,,
\label{bc}
\end{align}
$Q = n$ can be easily verified.
Notably, the energy of the magnetic skyrmion depends on $\gamma$, and it assumes the minimal value with $\gamma=\pi/2$. 
Further, rotationally invariant configuration \eqref{eq:ansatz} exists only for $n = 1$. 
On the contrary, for a special choice of potential, there are non-rotational solutions even for $n=1$. 
The authors of~\cite{Jaykka:2010bq} have found the broken rotationally symmetric solution of $n=1$ with the potential $V=(1-n_3)^2$ 
in terms of their energy minimization analysis. The Zeeman $V=(1-n_3)$ and the quadratic potential $V=(1-n_3)^2$
appear to be in rivalry in our model, suggesting that such deformation may appear. 
It seems a little outside the focus of the current study, therefore we maintain the 
symmetry to make the analysis simpler and also to make it easier to find compacton solutions.

Substituting (\ref{eq:ansatz}) with $n=1,\gamma=\pi/2$ into (\ref{eq:edens}), 
we define the energy density $\varepsilon [f]$
\begin{align}
    &\varepsilon[f]=\kappa_2 \left(f'^2+ \frac{\sin^2 f}{r^2} \right)
\nonumber \\
    &~~+\kappa_1 \left(f'+\frac{\sin 2f}{2r}\right)\sin \gamma 
	-\kappa_1\biggl(\cos f f'+\frac{\sin f}{r}\biggr)\sin \gamma
\nonumber \\ 
    &~~+\kappa_{4a} \frac{2\sin^2 ff'^2}{r^2}
    +\kappa_{4b}\left(f'^4 + \frac{2\sin^2 ff'^2}{r^2}+\frac{\sin^4 f}{r^4}\right)\nonumber\\
    &~~+\kappa_{0a}\left(1-\cos f\right)+\kappa_{0b}\left(1-\cos f\right)^2\,,
    \label{eq:edens-f}
\end{align}
where $f':=\dfrac{df(r)}{dr}$.
The function $f(r)$ satisfies the Euler equation, a nonlinear second-order ordinary differential equation.
\begin{align}
   &\kappa_2\left(rf''+f'-\frac{\sin 2f}{2r}\right)+\kappa_1 \sin^2 f \sin \gamma
\nonumber \\
   &+\kappa_{4a}\left(\frac{2\sin^2 f}{r} f''+\frac{\sin 2f}{r} f'^2 -\frac{2\sin^2 f}{r^2} f' \right)
\nonumber \\
   &+\kappa_{4b}\Biggl\{ \left( 6r f'^2+\frac{2\sin^2 f}{r} \right)f'' 
\nonumber \\
   &+\left( 2f'^2 +\frac{\sin 2f}{r}f' -\frac{2\sin^2 f}{r^2}\right)f' 
   -\frac{\sin 2f}{2r^2}+\frac{\sin 4f}{4r^2}\Biggr\}
\nonumber \\
   &-\frac{\kappa_{0a}}{2}r\sin f -\frac{\kappa_{0b}}{2}r\left(2\sin f -\sin 2f \right)=0\,.
   \label{Eulereq}
\end{align}

In the following part, we refer to the model in terms of its parameter 
settings: $[\kappa_2, \kappa_1 ,\kappa_{4a}, \kappa_{4b}, \kappa_{0a}, \kappa_{0b}]$.

\begin{figure*}[t]
  \begin{minipage}[b]{0.5\linewidth}
    \centering
    \includegraphics[keepaspectratio,scale=1.2,bb=0 0 193 137]{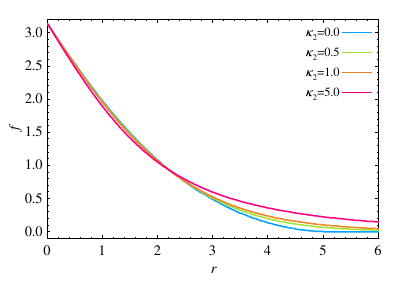}
  \end{minipage}\hspace{-0.5cm}
  \begin{minipage}[b]{0.5\linewidth}
    \centering
    \includegraphics[keepaspectratio,scale=1.2,bb=0 0 193 137]{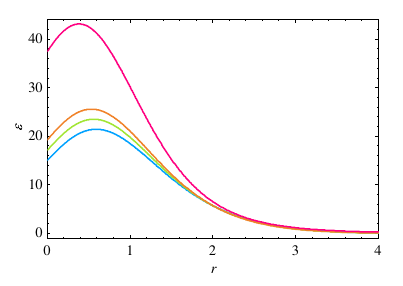}
  \end{minipage}

\caption{The skyrmions with $[\kappa_2, 0.0, 0.0, 1.0, 1.0, 1.0]$. The profile functions $f(r)$ (left) and the 
energy density $\varepsilon (r) $ (right). The model has the genuine-compacton solution for $\kappa_2=0.0$.}
  \label{fig:ho-pot}
\end{figure*}

\begin{figure}[t]
    \centering
    \includegraphics[keepaspectratio,scale=1.2,bb=0 0 193 137]{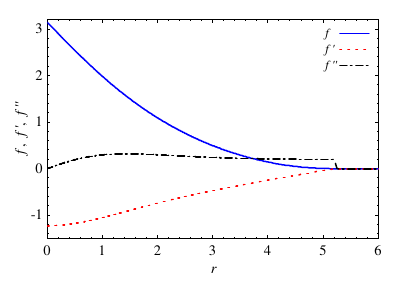}

\caption{We plot the compacton shown in Fig. \ref{fig:ho-pot}: 
the profile function and its derivatives $f(r),f'(r),f''(r)$~(the blue, red, and black lines), 
which clearly shows that the first derivative is continuous at the boundary $r = R = 5.231$.}
  \label{fig:ho-pot-dfdr}
\end{figure}

\begin{figure*}[t]
  \begin{minipage}[b]{0.5\linewidth}
    \centering
    \includegraphics[keepaspectratio,scale=1.2,bb=0 0 193 137]{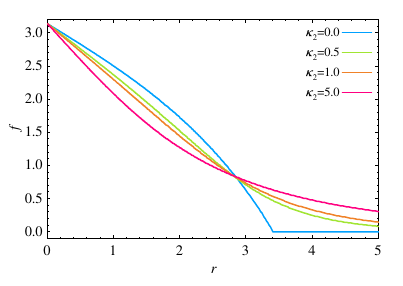}
  \end{minipage}\hspace{-0.5cm}
  \begin{minipage}[b]{0.5\linewidth}
    \centering
    \includegraphics[keepaspectratio,scale=1.2,bb=0 0 193 137]{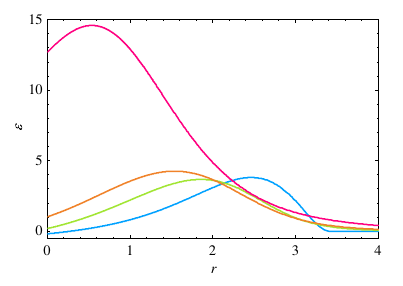}
  \end{minipage}

\caption{The skyrmions with $[\kappa_2, 1.0, 1.0, 0.0, 1.0, 0.0]$ of $\kappa_2 = 0.0, 0.5, 1.0, 5.0$. 
The profile functions $f(r)$ (left) and the energy density $\varepsilon (r)$ (right).
The restricted model $\kappa_2=0.0$ has the compacton solution (the blue line). }
  \label{fig:dm-sk-ze}
\end{figure*}

\begin{figure}[t]
    \centering
    \includegraphics[keepaspectratio,scale=1.2,bb=0 0 193 137]{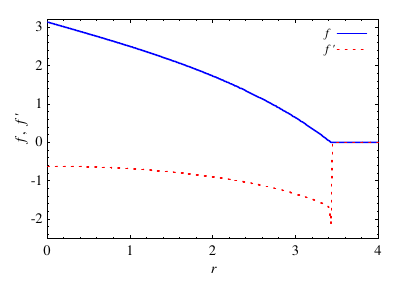}

\caption{We plot the compacton shown in Fig. \ref{fig:dm-sk-ze}: 
the profile function $f(r)$~(the blue line) and its derivative $f'(r)$ (the red line); this 
clearly shows that the derivative is not continuous at the 
boundary $r = R = 3.396$.}
  \label{fig:dm-sk-ze-dfdr}
\end{figure}

\section{\label{sec:3}Solutions: compactons emerging from the DMI term and the 4th- order terms}

\subsection{Models with the Skyrme term and the extended 4th term}

The model $[0, 0, \kappa_{4a}, 0, \kappa_{0a},0]$ 
is known as the restricted baby--Skyrme model that provides compacton solutions.
Compactons are solutions that reach their vacuum value $f\sim 0, (f'\sim 0)$ with finite radius $r = R$.  
Compacton is an advantageous form of the skyrmion lattice owing to its ability to smoothly 
connect the neighbors~\cite{Gudnason:2022aig}. 

We classify the compactons based on whether the 
function is continuously differentiable at the boundary: 
\begin{itemize}
\item \textrm{{\it Genuine}-compactons:}~~$f(R)=0,~\dfrac{df(r)}{dr}\biggr|_{r=R}=0$\,.
\item \textrm{{\it Weak}-compactons:}~~$f(R)=0,~\dfrac{df(r)}{dr}\biggr|_{r=R}\neq 0$\,.
\end{itemize}
For the weak-compacton case, even if the profile function is not differentiable, 
the energy density is still continuous because of the term $\sin^2f$. 
In addition, Speight~\cite{Speight:2010sy} also provides a classification approach in his paper for slightly 
different purpose. 

We consider a slightly generalized restricted model such as
$[0, 0, \kappa_{4a}, \kappa_{4b}, \kappa_{0a}, \kappa_{0b}]$. 

\subsubsection{$[0, 0, \kappa_{4a} ,0, \kappa_{0a}, \kappa_{0b}]$}
\label{sec:skyrme}

The model is the restricted baby--Skyrme model. 
Gisiger and Paranjape~\cite{Gisiger:1996vb} found the compacton in the model based on the Zeeman potential 
($\kappa_{0b} = 0$) by solving the Euler equations. Furthermore, Adam et al.~\cite{Adam:2010jr} found the compacton and 
non-compacton solutions in the models with different potential terms by solving the BPS equations. 
These potential terms~\cite{Adam:2010jr} are, for example, the Zeeman potential $V = (1-n_3)$, 
the new-baby potential $V = (1-n_3^2)$, and the squared Zeeman potential $V = (1-n_3)^2$. 
Here, we solve the Euler equations of the model based on two potential terms: the Zeeman and the squared Zeeman potential. 
We examine the mixed potential of the vacuum structure. From the boundary condition (\ref{bc}), the potential 
considers the minimum at $n_3 = 1$. We rewrite the potential as follows:
\begin{align}
    V[n_3]&=\kappa_{0a}(1-n_3)+\kappa_{0b}(1-n_3)^2 
\nonumber\\
&=(\kappa_{0a}+\kappa_{0b})(1-n_3)\biggl(1-\frac{\kappa_{0b}}{\kappa_{0a}+\kappa_{0b}}n_3\biggr) \label{eq:pot1}
\\
&= \kappa_{0b}\left(n_3-\frac{\kappa_{0a}+2\kappa_{0b}}{2\kappa_{0b}}\right)^2 - \frac{\kappa_{0a}^2}{4\kappa_{0b}} \label{eq:pot2},
\end{align}
where the parameters are set as $\kappa_{0a},\kappa_{0b}\neq 0$.
For (\ref{eq:pot1}), when the parameters are $\kappa_{0b}/(\kappa_{0a}+\kappa_{0b})=\pm 1$, 
i.e., $\kappa_{0a}=0$ or $\kappa_{0b}=-\kappa_{0a}/2$, these potentials become 
the squared Zeeman potential term or the new-baby potential term$V = (1-n_3^2)$. 
In the case of the squared Zeeman potential, the model has no compacton solutions. 
In the case of the new-baby potential, Adam et al. have solved the Bogomol'nyi equation and obtained the weak-compacton.
Here, we obtain the new compacton for $\kappa_{0a}\neq 0 $ and $\kappa_{0b}\neq -\kappa_{0a}/2 $.
According to (\ref{eq:pot2}), in the case of $\kappa_{0b}>0$, when the parameters 
satisfy $(\kappa_{0a}+2\kappa_{0b})/(2\kappa_{0b})\geq 1$, e.g., $\kappa_{0a}\geq 0$, the potential is always
positive and takes the minimum value: $V = 0$ at $n_3 = 1$.
Furthermore, for $\kappa_{0b}<0$, when the parameters are $(\kappa_{0a}+2\kappa_{0b})/(2\kappa_{0b})\leq 0$, e.g., 
$\kappa_{0a}\geq -2 \kappa_{0b}$, the potential is always positive and takes the minimum at $n_3 = 1$. 
As a result, in these conditions, the potential is suitable for determining the soliton solutions in the model. 
 
The solution found in \cite{Gisiger:1996vb} is apparently the weak-compacton case. It is directly verified 
by examining the analytical behavior at the compacton boundary $r = R$, 
where the profile function can be smoothly connected in vacuum.
We assume the series expansion around $r = R$.
\begin{align}
    f(r)=\sum_{s=0}^\infty A_s(R-r)^s\,.
\label{expansion}
\end{align}
The smoothness of the energy $f(R) = 0$ suggests that the expansion starts with $s > 0$. Here, it is sufficient
to consider the lowest-order term; thus, we substitute $f(r)\sim A_s (R-r)^s$ 
into the Euler equation and obtain the relation for the lowest-order contribution.
\begin{align}
    \frac{2\kappa_{4a}}{r}A_s^3 s (2s-1) (R-r)^{3s-2}-\frac{\kappa_{0a}}{2}r A_s (R-r)^s=0\,.
\end{align}
Obviously, it has a solution $s = 1$. 
This implies that there is a standard linear approach to vacuum, a typical feature for compactons in the restricted baby--Skyrme model.

For the case of the two potentials, the Zeeman and the squared Zeeman potential coexist, and
we can obtain the analytical solution in a similar manner. 
According to~\cite{Gisiger:1996vb}, we separate the equation \eqref{Eulereq} into
\begin{subnumcases}{}
\kappa_{4a} \biggl( 2 f''-\dfrac{2}{r}f'+2\cot f f'^2\biggr)
-\dfrac{\kappa_{0a}}{2}r^2\textrm{csc} f
\nonumber \\
\hspace{1cm}-\kappa_{0b}r^2(\textrm{csc} f-\cot f)  =0,\hspace{1cm}r\leq R,
\label{GPa} \\
\sin f=0,\hspace{4.5cm} r>R.
\label{GPb}
\end{subnumcases}
For simplicity, we employ the rescaling of the parameters as follows:
$\kappa_{0a}/\kappa_{4a}\to \kappa_{0a}, \kappa_{0b}/\kappa_{4a}\to \kappa_{0b}$. 
We define a new field: 
\begin{align}
  \mathcal{F}(r):=\cos f(r) -\frac{\kappa_{0a}+2\kappa_{0b}}{2\kappa_{0b}}
\end{align}
and the equation (\ref{GPa}) becomes a very simple form
\begin{align}
  \frac{d^2 \mathcal{F}}{dr^2}-\frac{1}{r}\frac{d\mathcal{F}}{dr}-\frac{\kappa_{0b}}{2}r^2\mathcal{F}=0\,.
\label{Eulereq3}
\end{align}
From the boundary condition (\ref{bc}), we have
\begin{align}
&\mathcal{F}(r=0)=-1-\frac{\kappa_{0a}+2\kappa_{0b}}{2\kappa_{0b}},
\nonumber \\
&\mathcal{F}(r=R)=1-\frac{\kappa_{0a}+2\kappa_{0b}}{2\kappa_{0b}}\,.
\end{align}
The equation (\ref{Eulereq3}) in $\kappa_{0b}>0$ can be solved analytically. 
The solution is
\begin{align}
&\mathcal{F}(r)=-\frac{\kappa_{0a}+4\kappa_{0b}}{2\kappa_{0b}}\cosh  
\left(\frac{\sqrt{\kappa_{0b}}r^2}{2\sqrt{2}}\right)
\nonumber \\
&\hspace{0.8cm}+\sqrt{\frac{2(\kappa_{0a}+2\kappa_{0b})}{\kappa_{0b}}} \sinh \left(\frac{\sqrt{\kappa_{0b}}r^2}{2\sqrt{2}}\right),\nonumber\\
& r\in \left[0,R=\frac{2^{3/4}}{\kappa_{0b}^{1/4}} \sqrt{\textrm{arccosh}\left( \frac{\kappa_{0a}+4\kappa_{0b}}{\kappa_{0a}}\right)}\right].
\end{align}

In $\kappa_{0b}<0$ and $\kappa_{0a}\geq 2|\kappa_{0b}|$, the solution is
\begin{align}
    &\mathcal{F}(r) = \frac{\kappa_{0a}-4\kappa_{0b}}{2\kappa_{0b}} \cos \left(\frac{\sqrt{\kappa_{0b}}r^2}{2\sqrt{2}}\right)\nonumber\\
    &\hspace{0.8cm}+\sqrt{\frac{2(\kappa_{0a}-2\kappa_{0b})}{\kappa_{0b}}}\sin \left(\frac{\sqrt{\kappa_{0b}}r^2}{2\sqrt{2}}\right),\nonumber\\
    &r\in \left[0,R=\frac{2^{3/4}}{\kappa_{0b}^{1/4}}\sqrt{\arccos \left(\frac{\kappa_{0a}-4\kappa_{0b}}{\kappa_{0a}}\right)}\right].
\end{align}

\begin{figure*}[t]
  \begin{minipage}[b]{0.5\linewidth}
    \centering
    \includegraphics[keepaspectratio,scale=1.2,bb=0 0 193 137]{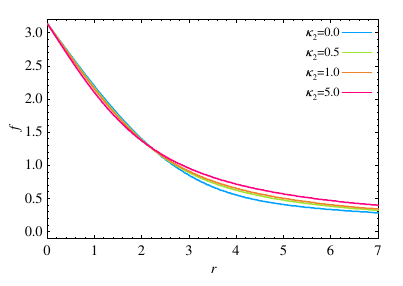}
  \end{minipage}\hspace{-0.5cm}
  \begin{minipage}[b]{0.5\linewidth}
    \centering
    \includegraphics[keepaspectratio,scale=1.2,bb=0 0 193 137]{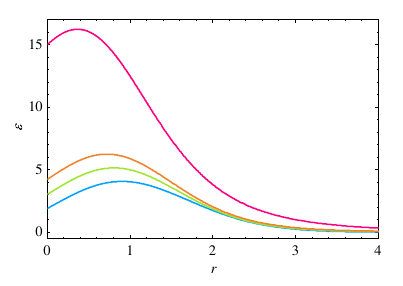}
  \end{minipage}

\caption{The skyrmions with $[\kappa_2, 1.0, 1.0, 0.0, 0.0, 1.0]$. The profile functions $f(r)$~(left) and the 
energy density $\varepsilon (r)$~(right). The restricted model $\kappa_2=0.0$ (the blue line) has no compacton. }
  \label{fig:sk-dm-sq}
\end{figure*}

\subsubsection{$[0, 0, 0, \kappa_{4b} ,\kappa_{0a}, \kappa_{0b}]$}	
\label{sec:quartic}

Interestingly, if we replace the Skyrme term with
the extended 4th term, we can obtain the genuine-compacton. 
First, we show that if there is a compacton in
the equation, we are essentially dealing with the genuine-compacton case. 
The following discussion is valid for the potentials: the Zeeman potential and the
mixtures of the Zeeman and the squared Zeeman potential. 
We assume at the boundary $f(R) = 0$, 
the Euler equation at the boundary $r = R$ becomes

\begin{align}
    2 f'(R)^2 \left\{3R  f''(R) +f'(R)\right\} =0\,,
\label{EeqaR}
\end{align}
has the solutions
\begin{align}
  \textrm{(i)}~f'(R)=0,~~\textrm{(ii)}~f''(R)=-\frac{f'(R)}{3R}.
\end{align}
First, we examine case (ii). If $f''(R) = -f'(R)/(3R)\neq 0$, the energy density is not continuous at $r = R$;
this is not what we aim for. If $f'(R) = 0$, the energy becomes continuous and subsequently becomes $f(R) = f'(R) = f''(R) = 0$.
It connects to the trivial vacuum solution $f(r)=0,r\in [0,R]$.
Therefore, case (i) $f'(R)=0$ should be employed for finding the nontrivial solutions in the genuine-compacton case. 
As a result, the second derivative $f''(r)$ is not continuous at $r = R$. 

This situation is again easy to confirm in terms of the expansion~(\ref{expansion}).
For the lowest order, we obtain
\begin{align}
    6 \kappa_{4b}r A_s^3 s^3 (s-1)r (R-r)^{3s-4}-\frac{\kappa_{0a}}{2}r A_s (R-r)^s=0\,.
\label{exeq2}
\end{align}
Here, we have a solution $s = 2$ for (\ref{exeq2}). 
This implies that a standard parabolic approach to the vacuum?a typical feature for the genuine-compacton case.

In this case, the analytical solution has not yet been found; thus, we numerically solve the Euler equation. 
We use the Newton-Raphson method with $N = 1000$ mesh points. 
We employ the standard rescaling scheme to the coordinate
\begin{align}
x=\frac{r}{1+r},~~x\in [0,1)\,.
\end{align}
The relative numerical errors of order $10^{-7}$.
Note that we always solve the Euler equation for the entire radial coordinate $x$ (not in the compact subset) 
even for the compactons, implying that compacton naturally arises in our numerical computation. 
We present our results for the Zeeman potential in Fig.\ref{fig:ho-pot}. As increasing $\kappa_2$, 
the tail of the profile function extends, and the maximum of the energy density is higher. 
This is because the kinetic term $\kappa_2 (\partial_i \bm{n} )^2>0$ exists in the energy density. 
Fig.\ref{fig:ho-pot-dfdr} shows the compacton solution and also the derivatives $f'(r), f''(r)$. 
This can easily be identified as the genuine-compacton case, and $f''(r)$ shows discontinuity around $r = R$.

It must be noted that the above discussion does not directly imply the
existence of compacton in the model. If we employ the squared Zeeman potential, 
the solution leads to normal skyrmions. 
This corresponds to the solution found in~\cite{Adam:2009px}, where the model is composed of the 
Skyrme term and the squared Zeeman potential.

\subsection{DMI model}\label{sec:DMI}

Here, we intend to find the DMI-mediated compactons of our model. 
First, according to the baby-skyrmions case, we set $\kappa_2 = 0$, i.e., the restricted model. 
In this case, we omit the Zeeman energy to obtain the solution. The parameter set is $[0,\kappa_1,0,0,0,\kappa_{0b}]$.
From~(\ref{Eulereq}), the equation becomes 
\begin{align}
    \kappa_1 \sin^2 f-\kappa_{0b}r\sin f(1-\cos f)=0\,.
\end{align}
For the nontrivial solutions $\sin f \neq 0$, except at the boundaries, 
\begin{align}
    \kappa_1 \sin f=\kappa_{0b}r(1-\cos f)\,.
\end{align}
By squaring on both sides, we obtain
\begin{align}
(\kappa_{0b}^2 r^2 +\kappa_{1}^2)\cos^2 f -2\kappa_{0b}^2r^2 \cos f +\kappa_{0b}^2 r^2    +\kappa_{1}^2=0\,.
\end{align}
We obtain the nontrivial solution of the form
\begin{align}
    \cos f =\frac{\kappa_{0b}^2r^2-\kappa_{1}^2}{\kappa_{0b}^2r^2+\kappa_1^2}\,.
\end{align}
This is exactly the solution found by Schroer in the equation of motion and also of a first-order 
Bogomol'nyi equation~\cite{Barton-Singer:2018dlh}. 
This solution is apparently not compacton. 

It can be confirmed by analysis of the expansion at the boundary (\ref{expansion}).
At the lowest order, we obtain
\begin{align}
    \kappa_1 A_s^2 (R-r)^{2s}-\frac{\kappa_{0a}}{2}r A_s (R-r)^s=0
\label{exeq3}
\end{align}
where the solution is $s=0$. 
This implies that the profile function is a constant at $r = R$, and it corresponds to 
the normal skyrmion solution. 

\begin{figure*}[t]
  \begin{minipage}[b]{0.5\linewidth}
    \centering
    \includegraphics[keepaspectratio,scale=1.2,bb=0 0 193 137]{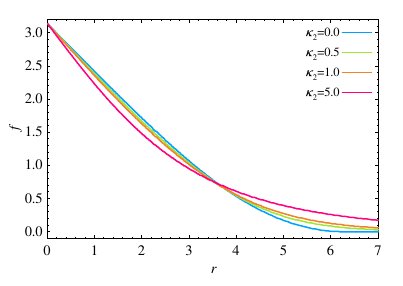}
  \end{minipage}\hspace{-0.5cm}
  \begin{minipage}[b]{0.5\linewidth}
    \centering
    \includegraphics[keepaspectratio,scale=1.2,bb=0 0 193 137]{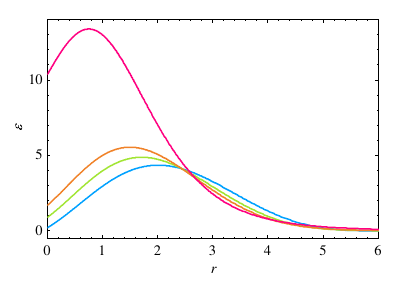}
  \end{minipage}

\caption{The skyrmions with $[\kappa_2, 1.0, 0.0, 1.0, 1.0, 0.0]$. 
The profile functions (left) and the energy density (right). The restricted model: $\kappa_2 = 0.0$ 
is the compacton solution (the blue line).}
  \label{fig:dm-4th-ze}
\end{figure*}

\begin{figure}[t]
    \centering
    \includegraphics[keepaspectratio,scale=1.2,bb=0 0 193 137]{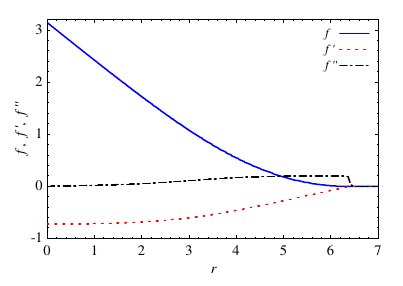}

\caption{We plot the compacton shown in Fig. \ref{fig:dm-4th-ze}: 
the profile function and the derivatives $f(r),f'(r)$~(the blue, the red lines) that clearly show that the derivatives are continuous at the boundary $r = R = 5.969$.}
  \label{fig:dm-4th-ze-dfdr}
\end{figure}

\begin{figure*}[t]
  \begin{minipage}[b]{0.5\linewidth}
    \centering
 \includegraphics[keepaspectratio,scale=0.3,bb=0 0 772 591]{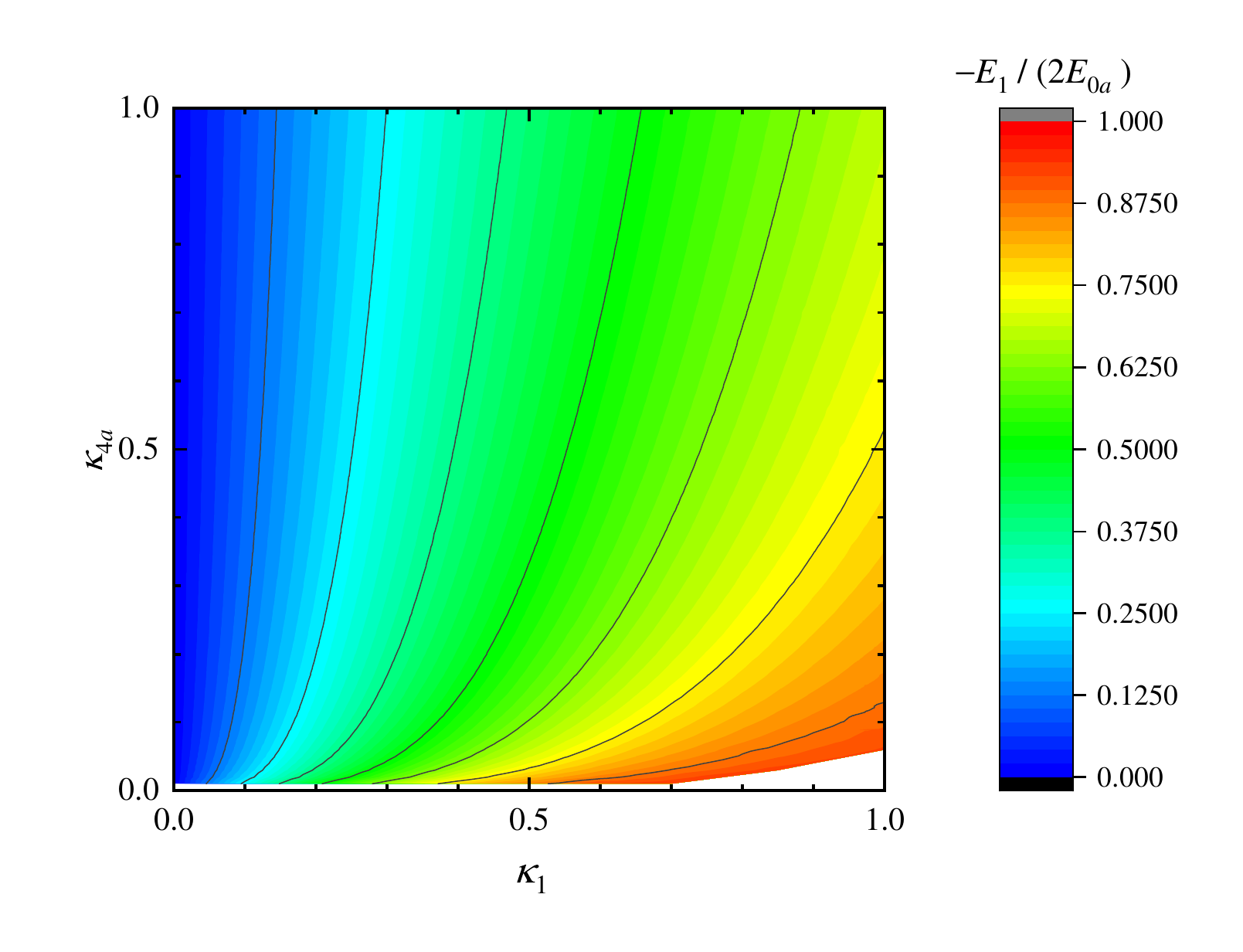}
  \end{minipage}\hspace{-0.5cm}
  \begin{minipage}[b]{0.5\linewidth}
    \centering
    \includegraphics[keepaspectratio,scale=0.3,bb=0 0 772 591]{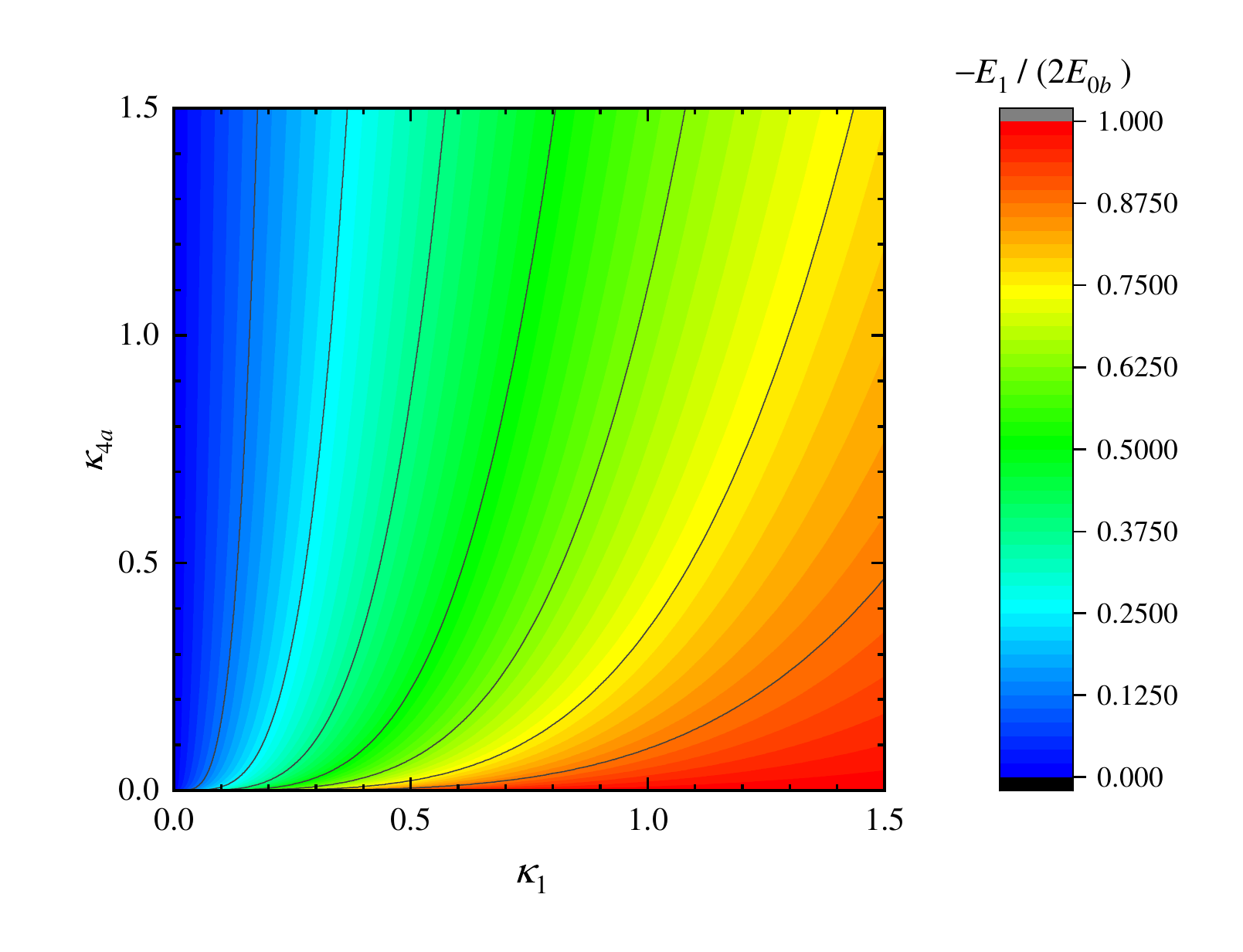}
  \end{minipage}

\caption{For the Derrick's theorem, we compute the ratio $-E_1/(2E_{0a})$ 
of the Skyrme - DMI -  Zeeman model:$[0.0, \kappa_1, \kappa_{4a} ,0.0 ,1.0 ,0.0]$~(left) 
and the ratio $-E_1/(2E_{0b})$ of the Skyrme - DMI - squared Zeeman model:
$[0.0, \kappa_1, \kappa_{4a}, 0.0, 0.0, 1.0]$~(right)
for various values of $(\kappa_1,\kappa_{4a})$.}
  \label{fig:sk-dm-ze-derrick}
\end{figure*}

\subsection{Inclusion of the DMI term and the 4th-order terms}\label{sec:DMI-quartic}

Next, to obtain a compacton solution, we add the 4th-order terms.  
The model is constructed based on the term in the restricted baby--Skyrme model. 
When both the DMI and the 4th-order terms are included, no analytical solutions 
exist, and we have to solve the equation numerically. 
We treat the model with the parameter set $[0,\kappa_1,\kappa_{4a},0,\kappa_{0a},\kappa_{0b}]$.
The model contains two types of derivative terms: DMI and the Skyrme term that hamper the scale invariance. 
The magnetic Skyrme model and the baby--Skyrme model possess soliton 
solutions for these terms and usually do not require a combination for stability. 
We shall look at Derrick's argument for the model. 
The energy applying the spatial rescaling $x\mapsto \mu x$ can be written as
\begin{align}
e(\mu)=E_2+\mu^{-1}E_1+\mu^2E_{4a}+\mu^{-2}(E_{0a}+E_{0b})\,. 
\end{align}
Taking the derivative with $\mu$, we obtain 
\begin{align}
&\frac{de(\mu)}{d\mu}\biggr|_{\mu=1}=-E_1+2E_{4a}-2(E_{0a}+E_{0b})=0\label{eq:e1e4}\,.
\end{align}
For evading the Derrick's argument, the potential energy should satisfy $E_{0a} + E_{0b} > 0$. 
For the parameters $\kappa_{0a}>0,\kappa_{0b}>0,$ or $\kappa_{0a}\geq -2\kappa_{0b},\kappa_{0b}<0$
 (\ref{sec:skyrme}). 
We divide both sides of (\ref{eq:e1e4}) by $E_{0a}+E_{0b}$, and we obtain
\begin{align}
    \frac{-E_1}{2(E_{0a}+E_{0b})}+\frac{E_4}{E_{0a}+E_{0b}}=1\,.
\end{align}
Since $-E_1,E_{4a}\geq 0$ 
\begin{align}
0\leq \frac{-E_1}{2(E_{0a}+E_{0b})},~~\frac{E_4}{E_{0a}+E_{0b}} \leq 1\,.
\end{align} 
In the magnetic Skyrme model, Derrick's theorem supports $-E_1/(2(E_{0a}+E_{0b}))=1$, while 
in the baby--Skyrme model, $E_{4a}/(E_{0a}+E_{0b})=1$. 
The ratio thus conveys the terms that have a dominant role in the stability of the solitons. We shall discuss this in the following sub-section. 

For the Skyrme term and the Zeeman potential, $[0, \kappa_1, \kappa_{4a}, 0, \kappa_{0a}, 0]$, 
the solutions are plotted in Fig.\ref{fig:dm-sk-ze} along with the non-compacton solutions $\kappa_2\neq 0$. 
Upon increasing $\kappa_2$, the tail of the profile function extends and changes the convex shape from upward to downward. 
The maximum energy density approaches the origin. 
This is because when the convex is downward, the range of $\pi/2<f\leq \pi$ reduces. 
At this time, the energy density of the kinetic term and the Skyrme term increases, whereas the contribution of the DMI term is negative. 
As a result, the energy density is enhanced in the vicinity of the origin. 
In Fig.\ref{fig:dm-sk-ze-dfdr}, we focus on the characteristic behavior of this solution;
it exhibits $f'(R)\neq 0$ at the boundary $r = R$ that appears similar to the compactons found by 
Gisiger and Paranjape~\cite{Gisiger:1996vb}. It is easy to verify how the feature is realized, as discussed below. 
If the boundary condition $f(R) = 0$ is substituted into the Euler equation (\ref{Eulereq}), 

\begin{align}
    \left.\frac{df(r)}{dr}\right|_{r=R} = \sqrt{\frac{\kappa_{0a}}{4\kappa_{4a}}}R\,.
\end{align}
at the boundary.
When the boundary is far from the origin $R\to \infty$, $df/dr\to \infty$, 
the energy density (\ref{eq:edens-f}) becomes divergent. 
Therefore, to avoid such singularity of the energy, the model has to choose a solution with a concrete finite radius, i.e., compacton.

When a different potential term is chosen, such as the squared Zeeman potential term, $[0, \kappa_1, \kappa_{4a}, 0, 0, \kappa_{0b}]$,
the solutions are not compactons (see Fig.\ref{fig:sk-dm-sq}). 
Upon increasing $\kappa_2$, the tail of the profile function extends and the maximum of energy density becomes higher and closer to the origin. 
Compared with Fig.\ref{fig:dm-sk-ze}, the change is moderate. 
If we assume $f(R) = 0$, the Euler equation at the boundary becomes 
\begin{align}
    \left.\frac{df(r)}{dr}\right|_{r=R}=0\,,
\end{align}
Nonetheless, this does not reveal anything regarding the compactness. 
In the next subsection, we present a new compacton solution satisfying this condition.

\subsection{Genuine - DMI - compacton case: $[0,\kappa_1,0,\kappa_{4b},\kappa_{0a},0]$}

Thus far, compactons emerged only in the restricted cases $\kappa_ 2 = 0$ 
, while for $\kappa_ 2\neq 0$, the solutions became normal skyrmions. 
Therefore, the kinetic term simply extends the tail of solutions.
For the 4th-order terms, the models with $\kappa_{4a}\neq 0 $ and $\kappa_{4b}=0$ 
include the weak-compacton case, while the models with $\kappa_{4a}=0$ and $\kappa_{4b}\neq 0$ 
include the genuine-compacton case~(see Fig.\ref{fig:dm-4th-ze} and also Fig.\ref{fig:dm-4th-ze-dfdr}). Therefore, the extended 4th- term has a role
in constructing the genuine-compacton case.
For our potentials, in the squared Zeeman potential case ($\kappa_{0a}=0$ and $\kappa_{0b}\neq 0$), 
the solutions are the baby-skyrmions, 
and in the Zeeman potential case ($\kappa_{0a}\neq 0$ $\kappa_{0b}=0$), the solutions become 
compactons.

We observed that the DMI is less effective for constructing the compactons. 
The reason is as follows: The DMI and potential terms do not have the derivatives in the Euler equation, and the major 
difference between the DMI and the potential terms are the dimensions; 
the potential is multiplied by $r$. 
The compacton radius $R$ is determined in terms of the behavior of 
the solutions at a large $r$, and apparently, the potential 
dominates rather than the DMI. That is also why the compactons are 
supported via potentials rather than the DMI.
From another perspective, we can easily confirm this based on the series expansion~(\ref{expansion}).
The condition of the leading order is as follows:
\begin{align}
&\kappa_1 A_s^2(R-r)^{2s}
+6\kappa_{4b}rA_s^3s^3(s-1)(R-r)^{3s-4}
\nonumber \\
&-\frac{\kappa_{0a}}{2}rA_s(R-r)^s=0\,.
\label{}
\end{align}
Therefore, the DMI term does not contribute to the lowest-order terms, and subsequently, 
the condition coincides with the one without the DMI term (\ref{exeq2}). 

In the next section, we shall examine the new model where the DMI plays an important role in compactons.

We consider the effect of the DMI and the Skyrme term concerning the stability (the existence) 
of the solutions from the perspective of Derrick's argument. 
We examine the value $-E_1/(2E_0)$ corresponding to the strength of several parameters for the models: 
$[0,\kappa_1, \kappa_{4a},0,\kappa_{0a},0]$ and $[0,\kappa_1, \kappa_{4a},0,0,\kappa_{0b}]$ 
If the solutions are obtained by DMI, it reaches 1, whereas it approximates 0, if obtained by the Skyrme term.
Fig.\ref{fig:sk-dm-ze-derrick} shows the result for the Zeeman and the squared Zeeman potential, respectively. 
These are reasonable results: 
for $\kappa_{4a}\to 0$, $-E_1/(2E_0)$ approaches $1$
and for $\kappa_1 \to 0$, $-E_1/(2E_0)$ approaches $0$.
Fig.\ref{fig:sk-dm-ze-derrick} shows no solutions in two regions: 
\begin{itemize}
\item[(i)] at $\kappa_{4a} \to 0$ for all $\kappa_1$, 
\item[(ii)] the lower right: $\kappa_1 \to 1$ with the small $\kappa_4$. 
\end{itemize}
(i) Without the Skyrme term, the restricted model of the Zeeman potential 
only has a half-skyrmion and no soliton solution.
(ii) Upon increasing $\kappa_{1}$, the model substantially moves to case (i). 
Thus, the blank grows as $\kappa_1$ increases.

In Fig.\ref{fig:sk-dm-ze-derrick}(right), the solution exhibits no such limiting behavior.

\begin{figure*}[t]
  \begin{minipage}[b]{0.5\linewidth}
    \centering
    \includegraphics[keepaspectratio,scale=1.2,bb=0 0 193 137]{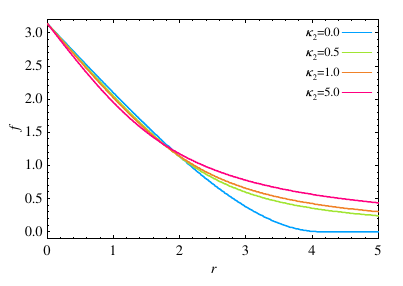}
  \end{minipage}\hspace{-0.5cm}
  \begin{minipage}[b]{0.5\linewidth}
    \centering
    \includegraphics[keepaspectratio,scale=1.2,bb=0 0 193 137]{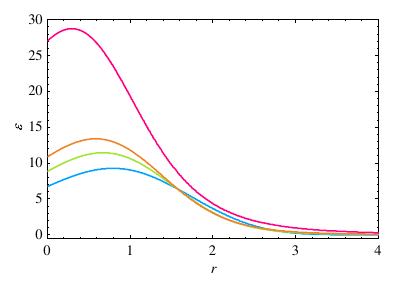}
  \end{minipage}

\caption{Skyrmions without potential~$[\kappa_2, 1.0, 1.0 ,0 ,0 ,0]$.  
The profile function $f(r)$ (left) and the energy density $\varepsilon (r)$ (right). 
The restricted model: $\kappa_2=0.0$ is the compacton solution (the blue line).}
  \label{fig:kin-dm-sk}
\end{figure*}

\begin{figure}[t]
    \centering
    \includegraphics[keepaspectratio,scale=1.2,bb=0 0 193 137]{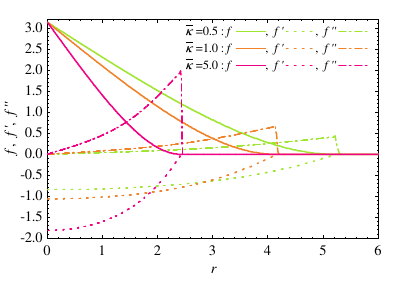}

\caption{We plot the compactons of~(\ref{restrictdm-sk}): 
the profile function and the derivatives $f(r),f'(r)$~(the blue, the red line), 
which clearly shows that the derivatives are continuous at the boundary $r = R = 5.231, 4.168, 2.431~(\bar{\kappa}=0.5, 1.0, 5.0)$.}
  \label{fig:mdm-sk_dfdr}
\end{figure}

\section{\label{sec:4}Solutions of the models without the potential terms}

In this paper, we have studied the normal models that always possess the potential terms. 
The kinetic term has no role in Derrick's theorem. In fact, the 
4th-order terms of the baby--Skyrme model and the DMI term of the 
magnetic Skyrme model along with the potential terms are 
responsible for the existence of the soliton solutions.

According to \cite{Ashcroft:2015jwa}, there is a new 
type of baby--Skyrme model without any potential term. 
The model is composed of the kinetic and Skyrme 
terms with the integer or fractional power of $\alpha,\beta$. 
The range of these parameters is examined to ensure stability 
with respect to rescaling. 

We propose a model that comprises the Skyrme and DMI without potential. 
The energy applying the spatial rescaling $x\mapsto \mu x$ can be written as
\begin{align}
e(\mu)=E_2+\mu^{-1}E_1+\mu^2E_{4a}\,.
\end{align}
There is no stationary point with $e(\mu)$ because $E_{4a}>0, E_1<0$ for $\gamma=\pi/2$. 
However, when we set $\gamma=-\pi/2$, it can take the extremum at
\begin{align}
\mu=\sqrt[3]{\frac{E_1}{2E_{4a}}},~~E_1,E_{4a}>0\,,
\end{align}
Accordingly, we may have a soliton solution to the model. 

We consider the model with $[\kappa_2, \kappa_1, \kappa_{4a} ,0 ,0 ,0]$. 
We present the results in Fig.\ref{fig:kin-dm-sk}. 
With an increase in $\kappa_2$, the tail of the solution extends, and the maximum of the energy density enhances at the origin 
because the gradient of the solution increases. 
In the case of $\kappa_2=0$, the solution becomes the compacton. 
For the restricted model ($\kappa_2 = 0$), the Euler equation is the following simple one-parameter equation.
\begin{align}
    2rf''-2f'+2r\cot f f'^2 -\bar{\kappa}r^2=0
\label{restrictdm-sk}
\end{align}
where $\bar{\kappa}:=\kappa_1/\kappa_{4a}$. 
Fig.\ref{fig:mdm-sk_dfdr} plots the $f(r),f'(r),f''(r)$ for several $\bar{\kappa}$. 
The $f,f'$ simultaneously becomes zero at $r = R$, where the $f''(R)$ remains finite that is likely genuine-compacton. Analytically, we can check this 
by substituting $f(R) = 0$ into (\ref{Eulereq}), we obtain $f'(R)=0$. 
In fact, it does not mean that there is a compacton solution in the model. 
However, we state that if a compacton exists, it should be the genuine-compacton case. With an increase in $\bar{\kappa}$, the compacton radius $R$ moves toward the origin. 

In this model, there is symmetry with respect to the inversion of the coefficient. 
Our model is (a)~$\gamma=-\pi/2, \kappa_{4a}>0$. The model
(b)~$\gamma=\pi/2, \kappa_{4a}<0,$ attains the same equation, 
where the energy density reverses the sign.
Here, we speculate whether the symmetry really exists. 
To confirm this, we add the kinetic term to the model, and (a)~provides the solution but not (b). 
The result of the heat-kernel expansion~(\ref{eq:edens}) shows that 
the kinetic, DMI, and Skyrme terms have the same sign. 
Furthermore, it is straightforward to verify that the model with $\kappa_2>0, \kappa_4<0$ is always unstable 
in the quantum stability analysis. Therefore, the above symmetry does not exist 
and is an artifact for the restricted model. 

In terms of the series expansion at the compacton boundary (\ref{expansion}), we have the condition
for the lowest order
\begin{align}
    \frac{2}{r}A_s^3 s (2s-1) (R-r)^{3s-2}-\bar{\kappa} A_s^2 (R-r)^{2s}=0\,,
\end{align}
that has the solution $s = 2$. 
This implies that there is a standard parabolic approach to vacuum for the genuine-compacton case.

\begin{figure*}[t]
  \begin{minipage}[b]{0.5\linewidth}
    \centering
    \includegraphics[keepaspectratio,scale=1.5,bb=0 0 193 137]{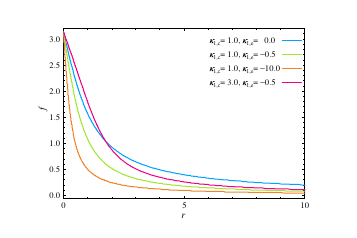}
  \end{minipage}\hspace{-0.5cm}
  \begin{minipage}[b]{0.5\linewidth}
    \centering
    \includegraphics[keepaspectratio,scale=1.5,bb=0 0 193 137]{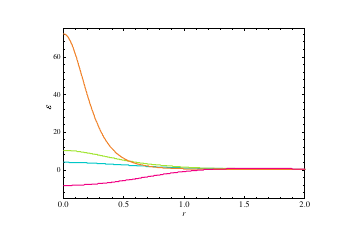}
  \end{minipage}

\caption{We plot the magnetic skyrmions of the model:$[1.0,\kappa_1(r),0,0,0,1.0]$ 
with the localized impurity $\kappa_1(r):=\kappa_{\textrm{1,c}}\exp(\kappa_{\textrm{1,e}}r)$. 
The profile function $f(r)$ (left) and the energy density $\varepsilon (r)$ (right). 
}
  \label{fig:impurity}
\end{figure*}

\begin{figure*}[t]
  \begin{minipage}[b]{0.5\linewidth}
    \centering
    \includegraphics[keepaspectratio,scale=1.5,bb=0 0 193 137]{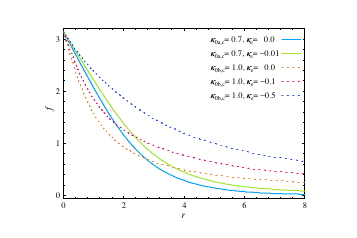}
  \end{minipage}\hspace{-0.5cm}
  \begin{minipage}[b]{0.5\linewidth}
    \centering
    \includegraphics[keepaspectratio,scale=1.5,bb=0 0 193 137]{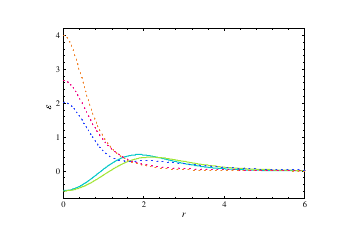}
  \end{minipage}

\caption{We plot the solutions with the impurity for the DMI and also the Zeeman potentials.
The solid line represents the results of the Zeeman potential with the Gaussian type impurity:
$[1.0,\kappa_1(r),0,0,\kappa_\textrm{0a}(r),0].~~\kappa_1(r)=1.0\times\exp[\kappa_\textrm{e}r^2],
\kappa_\textrm{0a}(r)=\kappa_\textrm{0a,c}\exp[2\kappa_\textrm{e}r^2]$.
The dashed line is the squared Zeeman potential with the exponential impurity:
$[1.0,\kappa_1(r),0,0,0,\kappa_\textrm{0b}(r)].~~
\kappa_1(r)=1.0\times\exp[\kappa_\textrm{e}r],\kappa_\textrm{0b}(r)=\kappa_\textrm{0b,c}\exp[2\kappa_\textrm{e}r]$.
The profile function $f(r)$ (left) and the energy density $\varepsilon (r)$ (right). 
}
  \label{fig:impurity2}
\end{figure*}

\section{\label{sec:5}Impurity model}

In this paper, we have used the ``constant'' gauge field, or the constant $D$. In \cite{Adam:2019yst},
the authors studied the supersymmetric extensions of a restricted baby--Skyrme model of the 
squared Zeeman potential with ``the impurity coupling. 
In particular, the analytical solution in the BPS equation is found for the exponentially localized 
impurity. 
Further, they claimed that the Bogomol'nyi equation of the baby--Skyrme model with impurity is identical to that of 
the magnetic skyrmions with the Dzyaloshinskii--Moriya interactions. This indicates that there is a certain relation
between the impurity and the DMI. Therefore, it is worth to investigate the present model with the DMI of 
the exponentially localized impurity. As we can directly see the effect, 
we concentrate on the model without the 4th order terms. 

We study the model
\begin{align}
[1.0,\kappa_1(r),0,0,0,1.0],~~\kappa_1(r):=\kappa_{\textrm{1,c}}\exp(\kappa_{\textrm{1,e}}r).
\end{align}
Here we employ the squared Zeeman potential as was in \cite{Adam:2019yst}.
Note that we could not find any solutions for the standard Zeeman potential. 
Fig.\ref{fig:impurity} presents several magnetic skyrmion solutions with the impurity. 
As increasing $|\kappa_\textrm{1,e}|$, the solutions shrink and the energy densities enhance. 
The existence of the solutions with large DMI coupling constant $\kappa_\textrm{1,c}$
is another notable effect of the impurity. It causes the energy of the solution to be a negative value.

A more challenging analysis is the case where the Zeeman potentials are also localizing.
If the model (\ref{eq:edens}) is obtained from the heat-kernel expansion of the fermionic 
vacuum functional~(\ref{vacuumfunctional}), the Zeeman potentials are also affected 
from the impurity. We study the two models
\begin{align}
&\textrm{the Zeeman:}~[1.0,\kappa_1(r),0,\kappa_\textrm{0a}(r),0],\hspace{5cm}
\nonumber \\
&\kappa_1(r)=1.0\times\exp(\kappa_{\textrm{e}}r^2),~
\kappa_\textrm{0a}(r)=\kappa_{\textrm{0a,c}}\exp(2\kappa_{\textrm{e}}r^2),
\\
&\textrm{the squared Zeeman:}~
[1.0,\kappa_1(r),0,0,\kappa_\textrm{0b}(r)],
\nonumber \\
&\kappa_1(r)=1.0\times\exp(\kappa_{\textrm{e}}r),~
\kappa_\textrm{0b}(r)=\kappa_{\textrm{0b,c}}\exp(2\kappa_{\textrm{e}}r).
\end{align}
In Fig.\ref{fig:impurity2}, we show some of our numerical results. The squared Zeeman potential and the exponential 
impurity support the existence of the solutions, but for the Zeeman potential, 
the Gaussian type impurity is required. The Zeeman potential permits a mild impurity, but 
the squared Zeeman potential allows a wider range of parameters. 
For both cases, as increasing $|\kappa_\textrm{1,e}|$, the solutions tend to dissipate. 
It is reasonable because when the impurity is more confined, the potentials soon decrease to zero at far from the origin. 
As a result, there is a critical point of upper limit for $|\kappa_\textrm{1,e}|$.  
For finding the exact value of such a point, a more thorough, extensive analysis seems to be required.

\section{Summary}

In this study, we have studied a generalization of the baby--Skyrme model with the inclusion 
of the Dzyaloshinskii--Moriya interaction (DMI). The model has been derived from 
the vacuum functional of fermions coupled with $O(3)$ nonlinear $\bm{n}$-fields and 
with a constant $SU(2)$ gauge background. We obtained
the effective action defined by the fermion determinant by integrating the fermionic fields. 
Based on the heat-kernel expansion for the determinant, we obtained the baby--Skyrme type model with the 
DMI and the two potential terms. 

In terms of the rotationally symmetric ansatz for $\bm{n}$-fields, we have obtained several normal
soliton solutions. For the restricted model, where the kinetic term is omitted, 
the compact skyrmions are obtained. The compactons are solutions with a 
finite radius, and the solutions encompass two cases of weak compacton and genuine compacton. 
These compactons are defined by a number of the differentiability at the boundary. 
The weak-compacton case is not continuously differentiable, and the genuine-compacton case is 
one-time differentiable. 
These are successfully obtained in terms of the choice of the 4th-order terms. 
The DMI has less effect in constructing compacton in this restricted model 
because the potential terms tend to dominate in the vicinity of the compacton radius. 
We proposed a new type of model for compactons without potential terms that comprise 
only the Skyrme term and the DMI term with opposite chirality. The solution is the 
genuine-compacton.

In the supersymmetric soliton models, the impurity coupling and the DMI are closely related. 
Therefore, using the exponential and the Gaussian functions of the DMI couplings, we 
investigated the effect of the localizing impurity. 

This study presents an initial step for the construction of soliton solutions for our new model. 
The following problems have to be solved in order:
\begin{itemize}
\item All our results were on the rotationally symmetric ansatz, and 
lifting this symmetry would be interesting. Especially, for the higher topological 
charges, non-rotationally symmetric solutions certainly exist. 
Also, even when $Q=1$, such deformation occurs with the quadratic Zeeman potential. 

\item Since the magnetic skyrmion often forms various platonic lattice structures, 
a novel structure based on the conjunction or competition between the DM and the 
4th-order terms may manifest in this model. 

\item The fermionic vacuum functional has its own soliton solutions for the model 
where the energy density comprises the sum of the valence fermions and an 
infinite sum of the Dirac sea fermions. The well-known Atiyah--Patodi--Singer 
index theorem implies the existence of such soliton solutions. 
The analysis is slightly complicated; nonetheless, the results are certainly interesting. 
\end{itemize}

We shall report on these issues in future studies.

\vskip 0.5cm\noindent
\begin{center}
{\bf Acknowledgment}
\end{center}

The authors would like to thank Pawe\l~Klimas
for his careful reading of this manuscript and the constructive feedback. 
We also appreciate Yuki Amari for his helpful advice regarding the constant gauge field~(\ref{DMgauge}).
We also thank Atsushi Nakamula, Yakov Shnir, Kouichi Toda, Shota Yanai for the fruitful discussions and comments. 
We also gratefully acknowledge the anonymous referee for the thoughtful and helpful comments.
N.S. is supported in part by JSPS KAKENHI Grant Number JP20K03278.

\appendix

\section{The heat-kernel expansion and the effective Skyrme--like model}

From the Dirac operator~(\ref{Diracop}), 
we define the Hamiltonian $h$ 
\begin{align}
&i\mathcal{D}=\sigma_3(-\partial_\tau-h),
\\
&h:=-\sigma_3\sigma_k\Bigl(\partial_k+iD\frac{\tau^k}{2}\Bigr)+\sigma_3 m \bm{\tau}\cdot\bm{n},~~k=1,2\,.
\end{align}
The baby--Skyrme model with the DMI emerges 
after subtracting the gauged (\ref{DMgauge}) vacuum state. 
We define the vacuum Hamiltonian with $\bm{n}_0=(0,0,1)$, 
\begin{align}
h_0=-\sigma_3\sigma_k\Bigl(\partial_k+iD\frac{\tau^k}{2}\Bigr)+\sigma_3 m\tau_3\,.
\label{hamiltonian0}
\end{align}
The choice for the gauge field (\ref{DMgauge}) and the vacuum Hamiltonian (\ref{hamiltonian0}) 
violate the $SU(2)$ symmetry of the theory. 

\renewcommand{\arraystretch}{2.0}
\begin{table*}[t]
  \caption{The results about the non-zero components of $\textrm{Tr}(b_n),~n\leqq 4$ with $O(D^k),~k\leqq 4$.}
  \centering
  \begin{tabular}{l|ll}
\hline\hline 
$b_2$: Tr$(\mathcal{V}^2)$  & (i)~$\textrm{O}(D^0):$&$ \frac{4m^2}{r^2} (\sin^2f+r^2f'^2)$
\\
&(ii)~$\textrm{O}(D^1):$&$-\frac{16m^2}{r}\sin\gamma\sin^2\frac{f}{2} (\sin f-rf')$
\\
&(iii)~$\textrm{O}(D^2):$&$4m^2\sin^2\frac{f}{2} (3-\cos f)$
\\
\hline\hline
$b_3$: Tr$\bigl(\mathcal{V}[\mathcal{H}_0,\mathcal{V}]
+2[\mathcal{H}_0,\mathcal{V}]\mathcal{V}\bigr)$ 
&(iv)~$\textrm{O}(D^2):$ & 
$-\frac{8m^2}{r^2}(1-\cos 2f+3r^2f'^2-r^2\cos 2ff'^2)$
\\
&(v)~$\textrm{O}(D^3):$ & 
$\frac{8m^2}{r}\sin\gamma \sin^2\frac{f}{2}(\sin f-rf')$
\\
\hline
~~~~~Tr$\bigl([\mathcal{H}_0,[\mathcal{H}_0,\mathcal{V}]]\bigr)$ 
&(vi)~$\textrm{O}(D^3):$ & 
$-\frac{8m^2}{r}\sin\gamma (\sin f+r\cos ff')$
\\
\hline\hline
$b_4$: Tr$(\mathcal{V}^4)$  
&(vii)~$\textrm{O}(D^0):$&$ \frac{4m^2}{r^4} (\sin^4f+6r^2\sin^2f f'^2+r^4f'^4)$  \\
&(viii)~$\textrm{O}(D^1):$ & $-\frac{32m^4}{r^3}\sin\gamma\sin^2\frac{f}{2} (\sin f-rf')^3$
\\
&(ix)~$\textrm{O}(D^2):$  & $\frac{16m^4}{r^2}\sin^2\frac{f}{2}\Bigl[(2-\cos 2\gamma)(3-\cos f)\sin^2f
+rf'\Bigl\{-16(2-\cos 2\gamma)\cos\frac{f}{2}\sin^3\frac{f}{2}$
\\
&&\hspace{1cm}$+r\bigl(6-2\cos f+\cos2\gamma(-1+\cos f)f'\bigr)\Bigr\}\Bigr]$
\\
&(x)~$\textrm{O}(D^3):$ & 
$\frac{32m^4}{r}\sin\gamma \sin^4\frac{f}{2}\Bigl\{-14\sin f+\sin 2f
-2r(-5+3\cos f)f'\Bigr\}$ 
\\
&(xi)~$\textrm{O}(D^4):$&$8m^4(35-28\cos f+\cos 2f)\sin^4\frac{f}{2}$ \\
\hline
~~~~~Tr$(\mathcal{V}[\mathcal{H}_0,[\mathcal{H}_0,\mathcal{V}]]$ 
&(xii)~$\textrm{O}(D^2):$ 
&$\frac{64m^4}{r^2}\Bigl[2\cos^2\gamma\sin^2f+rf'\Bigl\{2\cos^2\gamma\sin 2f+r(1+\cos 2\gamma\cos^2f)f'\Bigr\}\Bigr]$
\\
$\hspace{0.5cm}+3[\mathcal{H}_0,[\mathcal{H}_0,\mathcal{V}]]\mathcal{V})$
&(xiii)~$\textrm{O}(D^3):$  & $128 m^4\sin\gamma\sin^2ff'$
\\
&(xiv)~$\textrm{O}(D^4):$&$\frac{16m^2}{r^2}\Bigl\{(4m^2r^2+\cos^2\gamma)\sin^2f
-rf'(\sin^2\gamma\sin 2f-4\cos^2\gamma\cos^2ff')\Bigr\}$ \\
\hline 
~~~~~3Tr$([\mathcal{H}_0,\mathcal{V}]^2)$ 
&(xv)~$\textrm{O}(D^2):$ &
$-\frac{48m^4}{r^2}\Bigl\{2\cos^2\gamma\sin^2f+2r\cos^2\gamma\sin 2ff'
+r^2\bigl(1+\cos 2\gamma\cos^2f\bigr) f'^2\Bigr\}$ 
\\
&(xvi)~$\textrm{O}(D^3):$ & $-96m^4\sin\gamma\sin^2ff'$
\\
&(xvii)~$\textrm{O}(D^4):$&$\frac{12m^2}{r^2}\Bigl\{(1-4m^2r^2)\sin^2f+r^2f'^2)\Bigr\}$
\\
\hline\hline 
  \end{tabular}
\end{table*}
\renewcommand{\arraystretch}{1.0}

Here, the (3+1)-QCD effective model~\cite{Dhar:1985gh,Ebert:1985kz, Reinhardt:1989st,Wakamatsu:1990ud} is similarly analyzed, 
where a regularized action must be introduced because the action is generally divergent. 
According to \cite{Ebert:1985kz, Reinhardt:1989st}, we define the suitable-time-regularized action expressed as follows:
\begin{align}
\omega_R(\bm{n})\to -\frac{1}{2}\int_{1/\Lambda^2}ds s^{-1}\textrm{Tr}\exp(-s\mathcal{D}^\dagger\mathcal{D}) \,.
\end{align}
A substantial difference between the (2+1)- and the (3+1)-models is noted. Because 
in the (2+1)-model, (\ref{efactionr}) becomes finite after suitably subtracting the vacuum contribution, 
and the ultraviolet cutoff need not be introduced. 
When we consider the Dirac sea contribution to the total energy, 
the cutoff significantly improves the numerical convergence; thus, we retain it in the 
formulation. In this study, we examine the resulting Skyrme models found by this expansion; accordingly, 
we set $\Lambda\to\infty$. The energy is expressed as follows: 
$\omega_R(\bm{n})=-\int_0^\infty d\tau E_0$, and 
\begin{align}
E_0=\frac{1}{4\sqrt{\pi}}\int_{1/\Lambda^2}^\infty ds s^{-3/2}\textrm{Tr} K(s),~~K(s):=\exp(-sh^2)\,.
\end{align}
For the heat-kernel expansion, the proper-time kernel is expressed as
\begin{align}
\mathcal{H}=\mathcal{H}_0+\mathcal{V}\,,~~\mathcal{H}:=h^2\,,~~\mathcal{H}_0:=h_0^2\,,
\end{align}
and 
\begin{align}
K(s):=K_0(s)K_1(s),
\end{align} 
where
\begin{align}
K_0(s)=\exp(-s\mathcal{H}_0)\,,
\end{align}
and the interaction part is
\begin{align}
K_1(s)=\textrm{T}\exp\biggl[-\int_0^s ds'K_0(-s')\mathcal{V}K_0(s')\biggr]\,,
\end{align}
where T denotes the proper-time ordering. 
The interaction part satisfies the differential equation: 
\begin{align}
[\partial_s+K_0(-s)\mathcal{V}K_0(s)]K_1(s)=0,~~K_1(s=0)=1\,.
\label{heatequation}
\end{align}
It has the heat expansion
\begin{align}
K_1(s)=\sum_{n=0}^\infty s^nb_n,~~b_0=1\,.
\label{heatexpansion}
\end{align}
The heat coefficients $b_n$ can be easily obtained by plugging (\ref{heatexpansion}) into 
(\ref{heatequation}); the first a few terms are 
\begin{align}
&b_1=-\mathcal{V},~~
\nonumber \\
&2b_2=\mathcal{V}^2-[\mathcal{H}_0,\mathcal{V}],
\nonumber \\
&6b_3=-\mathcal{V}^3+\bigl(\mathcal{V}[\mathcal{H}_0,\mathcal{V}]
+2[\mathcal{H}_0,\mathcal{V}]\mathcal{V}\bigr)-[\mathcal{H}_0,[\mathcal{H}_0,\mathcal{V}]],
\nonumber \\
&24b_4=\mathcal{V}^4-\bigl(\mathcal{V}^2[\mathcal{H}_0,\mathcal{V}]
+2\mathcal{V}[\mathcal{H}_0,\mathcal{V}]\mathcal{V}
+3[\mathcal{H}_0,\mathcal{V}]\mathcal{V}^2\bigr)
\nonumber \\
&+\bigl(\mathcal{V}[\mathcal{H}_0,[\mathcal{H}_0,\mathcal{V}]]
+3[\mathcal{H}_0,[\mathcal{H}_0,\mathcal{V}]]\mathcal{V}\bigr)
+3[\mathcal{H}_0,\mathcal{V}]^2
\nonumber \\
&-[\mathcal{H}_0,[\mathcal{H}_0,[\mathcal{H}_0,\mathcal{V}]]]\,.
\label{coefficient}
\end{align}
The energy in the heat-kernel expansion is
\begin{align}
E_0=\frac{1}{4\sqrt{\pi}}\int_{1/\Lambda^2}^\infty ds s^{-3/2}\sum_{n=0}^\infty \textrm{Tr} (K_0(s)b_n)\,.
\end{align}
For evaluating the trace Tr, it considers the Lorentz, flavor (isospin), and also 
plain wave. 
\begin{align}
h_{\rm plain}|\phi_\nu^0\rangle =\epsilon_\nu^0|\phi_\nu^0\rangle\,,~~
h_{\rm plain}=-\sigma_3\sigma_k\partial_k+\sigma_3 m\,,
\end{align}
the energy $E_0$ becomes
\begin{align}
E_0=\frac{1}{2}\sum_{n=0}^\infty\sum_\nu |\epsilon_\nu^0|^{1-2n}\Gamma\biggl(n-\frac{1}{2},\Bigl(\frac{\epsilon_\nu^0}{\Lambda}\Bigr)^2\biggr)
\langle\phi_\nu^0|b_n|\phi_\nu^0\rangle\,.
\end{align}
The explicit form of $\mathcal{V}$ is
\begin{align}
\mathcal{V}&=m\sigma_k\bigl(\bm{\tau}\cdot\partial_k\bm{n}-i[\bm{A}_k,\bm{\tau}\cdot\bm{n}-\tau_3]\bigr)
\nonumber \\
&=m\sigma_k\bigl\{\bm{\tau}\cdot\partial_k\bm{n}
+D\bigl((\bm{\tau}\times\bm{n})_k-\epsilon_{k\ell}\tau_\ell\bigr)\bigr\}\,,~~k,\ell=1,2
\end{align}
where the $\epsilon_{k\ell}$ is the antisymmetric tensor. 
The first nonzero contribution to the energy is the second-order term: $n = 2$
\begin{align}
&E_0^{(n=2)}=\kappa_2\int d^2x \Bigl\{(\partial_i\bm{n})^2+2D\bm{n}\cdot(\nabla\times\bm{n})
\nonumber \\
&\hspace{1cm}-2D(\partial_1n_2-\partial_2n_1)
\nonumber \\
&\hspace{1cm}+2D^2(1-n_3)+D^2(1-n_3)^2
\Bigr\}\,,
\label{E2}
\\
&\kappa_2:=\frac{m}{8\pi^{3/2}}\Gamma\biggl(\frac{1}{2},\Bigl(\frac{m}{\Lambda}\Bigr)^2\biggr)
\underset{\Lambda\to\infty}{\longrightarrow} \frac{m}{8\pi}\,.
\end{align}
If we set $D = 0$ for $n\geq 3$ and restrict ourselves to the case of no DMI to the Skyrme or higher-order corrections
the results of the subsequent order $n = 4$ is relatively easy to obtain
\begin{align}
&E_0^{(n=4)}=\kappa_4\int d^2x \Bigl\{2(\partial_i\bm{n}\times\partial_j\bm{n})^2
+(\partial_i\bm{n})^2(\partial_j\bm{n})^2\Bigr\}\,, \\
&\kappa_4:=\frac{1}{96\pi^{3/2}m}\Gamma\biggl(\frac{5}{2},\Bigl(\frac{m}{\Lambda}\Bigr)^2\biggr)
\underset{\Lambda\to\infty}{\longrightarrow}
\frac{1}{128\pi m}\,.
\label{E4}
\end{align}

For full calculation up to the 4th-order contribution is almost straightforward but the 
results are cumbersome. 
For the actual calculation, the following relations may be useful:
\begin{align}
&\mathcal{H}_0:=\mathcal{H}_0^{d^2}+\mathcal{H}_0^{d^1}+\mathcal{H}_0^M\,,
\nonumber \\
&\mathcal{H}_0^{d^2}:=-\partial_k^2,~~\mathcal{H}_0^{d}:=iD(\tau_1\partial_1+\tau_2\partial_2)\,,
\nonumber \\
&\mathcal{H}_0^M:=\frac{D^2}{2}(1-\sigma_3\tau_3)+Dm(\sigma_1\tau_2-\sigma_2\tau_1)+m^2\,;
\nonumber 
\end{align}
\begin{align}
&[\mathcal{H}_0^{d^2},\mathcal{V}]\equiv \sum_{i=1}^2\Phi_i\partial_i\,,
\nonumber \\
&\Phi_i:=-2mD(\sigma_1\tau_2\partial_i n_3-\sigma_1\tau_3\partial_i n_2-\sigma_2\tau_1\partial_i n_3+\sigma_2\tau_3 \partial_i n_1)\,,
\nonumber \\
&[\mathcal{H}_0^{d^1},\mathcal{V}]_i\equiv \Psi^{(0)}_i+\Psi^{(1)}_i\partial_i\,,
\nonumber \\
&\Psi^{(0)}_1=mD^2(-\sigma_1\tau_2\partial_1n_1+\sigma_1\tau_2\partial_1n_2+\sigma_1\tau_3\partial_1n_3+i\sigma_2\partial_1n_3)\,,
\nonumber \\
&\Psi^{(1)}_1=2mD^2(-\sigma_2\tau_2n_1+\sigma_1\tau_2n_2+\sigma_1\tau_3(n_3-1))\,,
\nonumber \\
&\Psi^{(0)}_2=mD^2(\sigma_2\tau_1\partial_2n_1-\sigma_1\tau_1\partial_2n_2+\sigma_2\tau_3\partial_2n_3-i\sigma_1\partial_2n_3)\,,
\nonumber \\
&\Psi^{(1)}_2=2mD^2(\sigma_2\tau_1n_1-\sigma_1\tau_1n_2+\sigma_2\tau_3(n_3-1))\,,
\nonumber 
\end{align}
In Table I, we summarize all the terms $\textrm{Tr}(b_i)$ within the rotationally symmetric ansatz (42).

In terms of Table I, we can define the energy density of the full model.  We present for each $n$ 
\begin{align}
&\varepsilon^{(n=2)}=\kappa_2 \biggl[\frac{1}{r^2}(\sin^2f+r^2f'^2)-D\frac{4}{r}\sin^2\frac{f}{2}(\sin f-rf')
\nonumber \\
&+D^22(3-\cos f)\sin^2\frac{f}{2}\biggr]\,,
\label{edens2}
\end{align}
\begin{align}
&\varepsilon^{(n=3)}=\kappa_3 \biggl[D^2\frac{4}{3mr^2}(-1+\cos 2f-3r^2f'^2+r^2\cos 2f f'^2)
\nonumber \\
&+D^3\frac{4}{3mr}\biggl\{
\sin^2\frac{f}{2}(\sin f-rf')+\sin f+r\cos ff'
\biggr\}
\biggr]\,,
\label{edens3}
\end{align}
\begin{align}
&\varepsilon^{(n=4)}=\kappa_4 \biggl[
\frac{1}{r^4}(\sin^4f+6r^2\sin^2ff'^2+r^4f'^4)
\nonumber \\
&-D\frac{8}{r^3}(\sin f-rf')^3
-D^2\frac{12}{r^2}\biggl\{(-3+\cos f)\sin^2f
\nonumber \\
&+rf'(4\sin f-2\sin 2f+r(-3+\cos f)f'\biggr\}
\nonumber \\
&+D^3\frac{2}{r}\sin^2\frac{f}{2}\biggl\{29\sin f-16\sin 2f +\sin 3f
\nonumber \\
&-2r \bigl(17-12 \cos f+3\cos 2f\bigr)f'\biggr\}
\nonumber \\
&+D^4\biggl\{2(35-28\cos f+\cos 2f)\sin^4\frac{f}{2}
\nonumber \\
&+\Bigl(\frac{3}{m^2r^2}+4\Bigr)\sin^2f-\frac{4}{m^2r}\sin 2ff'+\frac{3}{m^2}f'^2\biggr\}
\biggr]\,,
\label{edens4}
\end{align}
where the coupling constants
\begin{align}
&\kappa_2=\frac{m}{8\pi},~~\kappa_3=\frac{1}{32\pi},~~
\kappa_4=\frac{1}{128\pi m}.
\label{coupling}
\end{align} 

Now, we discuss how our model is built from (\ref{edens2})-(\ref{edens4}).
Physically, the constant $D$ is supposed to be a  small value, thus we omit the 
terms of O$(D^k),k\geqq 2$, except for the Zeeman potentials.
Another possibility is to simply include the potentials as external terms and 
leave out the Zeeman potentials ((iii) in Table I, or the last term of (\ref{edens2})) for the sake of consistency. 

The heat-kernel expansion is justified for $m>1$, as can be seen from (\ref{coupling}).
In a slightly different context, the normalizable zeromodes of the fermion coupled with the baby-skyrmion
was studied in~\cite{Amari:2019tgs}. The modes are emerged above some critical value of $m$ (in \cite{Amari:2019tgs}, 
we showed the plot of the level crossing of the Dirac fermions in the case of $m=1$. 
For smaller $m$, no crossing occurs.) 
As a result, the coupling constants apparently satisfy the relation $\kappa_2>\kappa_3>\kappa_4$ in $m>1$.  
For example, $\kappa_2\sim 0.080, \kappa_4\sim 0.0012$ in $m=2$.  
Therefore, in this paper, we decided to omit the DMI-mediated term 
in $\varepsilon^{(n=4)}$ ((viii) in Table I, or the second term of (\ref{edens4})).
(There may be some small effects on the results within a numerical modeling such as $\kappa_4\sim  1$.)
As a result, in its simplest form, the model can be reduced in (\ref{eq:edens-f}).

We can fix the coupling constant of (\ref{eq:edens-f}), such as
\begin{align}
&\kappa_1:=2\kappa_2D,~~\kappa_{4a}=2\kappa_4,~~\kappa_{4b}:=\kappa_4\,,
\nonumber \\
&\kappa_{0a}:=2\kappa_2D^2,~~\kappa_{0b}:=\kappa_2D^2\,.
\label{eq:kappa}
\end{align}
In this study, we do not restrict our analysis to the above relations (\ref{eq:kappa}). In fact, 
we freely choose these parameters to determine the range of potential solutions.  

\bibliography{restskyrme}

\end{document}